\documentclass[10pt,conference,a4paper]{IEEEtran}
\usepackage[T1]{fontenc}
\usepackage{amsmath}
\usepackage[cmintegrals]{newtxmath}
\usepackage{url}
\usepackage{soul,color,graphicx,float,epstopdf}
\usepackage{tablefootnote,textcomp,gensymb}
\usepackage{eurosym,booktabs,multirow}
\usepackage[caption=false]{subfig}
\usepackage{amsfonts} 
\usepackage{algorithmic}
\usepackage{array}
\usepackage{stfloats}
\usepackage{float}
\usepackage{mathtools}
\usepackage{graphicx}
\usepackage{pdfpages}
\usepackage{subfig} 
\usepackage{balance}
\usepackage{xcolor}
\usepackage{comment}
\usepackage[normalem]{ulem}

\usepackage{tikz}
\usetikzlibrary{arrows,shapes,positioning,shadows,trees}

\tikzset{
  basic/.style  = {draw, text width=4cm, drop shadow, font=\sffamily, rectangle},
  root/.style   = {basic, rounded corners=2pt, thin, align=center,
                   fill=red!20},
  level 2/.style = {basic, rounded corners=6pt, thin,align=center, fill=blue!20,
                   text width=9em},
  level 3/.style = {basic, thin, align=left, fill=pink!20, text width=8em}
}

\definecolor{Orange}{rgb}{1,0.5,0}
\newcommand{\todo}[1]{\textsf{\textbf{\textcolor{Orange}{[#1]}}}}

\begin{document}
\title{Variational Autoencoder Assisted Neural Network Likelihood RSRP Prediction Model}

\author{\IEEEauthorblockN{Peizheng Li\IEEEauthorrefmark{1},
Xiaoyang Wang\IEEEauthorrefmark{1},
Robert Piechocki\IEEEauthorrefmark{1},
Shipra Kapoor \IEEEauthorrefmark{2},
Angela Doufexi\IEEEauthorrefmark{1},
Arjun Parekh \IEEEauthorrefmark{2}
}\\ 
\IEEEauthorblockA{\IEEEauthorrefmark{1} University of Bristol, UK\\ 
\IEEEauthorrefmark{2} Applied  Research,  BT,  UK\\
Email: {\{peizheng.li, xiaoyang.wang, A.Doufexi,  R.J.Piechocki\}@bristol.ac.uk}\\
{\{shipra.kapoor, arjun.parekh\}@bt.com}}}
\maketitle
\begin{tikzpicture}[remember picture, overlay]
  \node[minimum
  width=4in,fill=white!100,text=black,font=\normalsize, align=left] at ([yshift=-1cm, xshift=0cm]current page.north)  {This paper has been accepted for publication in PIMRC 2022, but has not been fully edited. Content may change prior to final publication.};
\end{tikzpicture}
\begin{abstract}

Measuring customer experience on mobile data is of utmost importance for global mobile operators. The reference signal received power (RSRP) is one of the important indicators for current mobile network management, evaluation and monitoring. Radio data gathered through the minimization of drive test (MDT), a 3GPP standard technique, is commonly used for radio network analysis. Collecting MDT data in different geographical areas is inefficient and constrained by the terrain conditions and user presence, hence is not an adequate technique for dynamic radio environments. In this paper, we study a generative model for RSRP prediction, exploiting MDT data and a digital twin (DT), and propose a data-driven, two-tier neural network (NN) model. In the first tier, environmental information related to user equipment (UE), base stations (BS) and network key performance indicators (KPI) are extracted through a variational autoencoder (VAE). The second tier is designed as a likelihood model. Here, the environmental features and real MDT data features are adopted, formulating an integrated training process. On validation, our proposed model that uses real-world data demonstrates an accuracy improvement of about 20\% or more compared with the empirical model and about 10\% when compared with a fully connected prediction network.


\end{abstract}

\begin{IEEEkeywords}
RSRP, variational autoencoder, likelihood model, neural network
\end{IEEEkeywords}
\section{Introduction}
\label{sec:intro}
The advent of the fifth-generation (5G) network provides a unique potential for UEs through three application themes namely enhanced mobile broadband (eMBB), ultra-reliable low-latency communications (URLLC) and massive machine-type communications (mMTC). These network advancements bring challenges to the complex heterogeneous radio access networks (RAN) and ultra-dense networks (UDN) such as BS deployment, radio coverage and capacity planning. Mobile network providers rely on a series of key performance indicators (KPIs) to understand, analyse and assess network performance for coverage and capacity. One of the KPIs for coverage analysis is signal strength - reference signal received power (RSRP). To measure RSRP values as well as other network metrics drive tests are usually performed.  It requires significant human efforts, explicit hardware and substantial capital expenditure (CAPEX). Moreover, the measurements recorded via this method are limited. It only reflects network performance for a short period for a particular location that lacks comprehensive spatial-temporal data collection and assessment. To overcome the challenges for RSRP measurement using the drive test, the 3rd Generation Partnership Project (3GPP) in Release 9 introduced the Minimisation of Drive test (MDT) methodology. Here, radio measurements are collected using an individual's mobile device that is logged into the network. Each UE feedback its network experience to the associated BS. 

Generally, there are two approaches for predicting RSRP. The first is via the 3GPP standard-based empirical model. Based on field measurements from different terrain and scenarios, such as urban, rural, suburban, macro or microcells, it summarizes the propagation rules based on a large number of test values. Typical empirical-based path loss models as documented in 3GPP TR 38.901 \cite{3GPP_release}. Such modelling depicts the channel properties in a general and coarse way, which may not be accurate enough for specific environments. The second method is the data-driven approach. Instead of modelling the propagation model, the correlation between environmental features in an area and the corresponding RSRP values are directly estimated. This approach needs a large amount of historical RSRP data that could be obtained through MDT. However, data collected by MDT has the following issues (a) the signal strength between UE devices at the same location and time could differ more than $\pm$6dB \cite{MDT_test}. (b) Inaccurate location information for indoor UEs. (c) Some locations only contain limited data points due to imbalanced UE distribution. Estimating and predicting RSRP values from limited and inaccurate measurements leads to ineffective Quality of Service (QoS) analysis.
Recently, efforts towards employing Artificial intelligence (AI) or machine learning (ML) techniques in RAN are progressing swiftly. One example is the introduction of the Radio Intelligence controller (RIC) module in the radio network architecture of Open RAN (O-RAN) \cite{li2021rlops}. 
ML-enabled algorithms provide feasible and accurate solutions to use cases such as intelligent coverage analysis, smart handover, load balancing, etc. In this work, we propose to use an ML model for RSRP prediction, which assists in assessing network coverage in the focused area.

This paper presents a generative model for accurate RSRP prediction based on a well-designed neural network (NN) architecture. We not only utilise the historical real data of RSRP but also consider the geographical statistics information. The correlation between geographical information and RSRP distributions is then mapped through data compression. Regarding the extraction of environmental features, we construct images that can reflect the transmission environment from BS to UEs, using them as auxiliary training features of the RSRP prediction model. The BS-UE association modelling is accomplished by a modified digital twin (DT), while the process of feature extraction is completed by a variational autoencoder (VAE), with a convolutional neural network (CNN) as a backbone. Once the model is trained, the encoder of VAE serves as the environment feature extractor in this paper. The low-dimensional latent variables will be used as the environmental features to assist the RSRP prediction. 

In regards to the RSRP prediction model, a multi-layer perception (MLP) trained in a supervised learning manner is applied. Due to changes in the transmission environment, the RSRP value recorded at each location is time-varying. From a statistical point of view, the RSRP values recorded at this location conform to a normal distribution. To estimate the normal distribution, the MLP model is designed as a likelihood model. It takes the output of the aforementioned encoder and BS-recorded features as inputs and outputs the mean and variance. The main contributions of this paper are summarised below:
\begin{itemize}
    \item We propose a likelihood NN model for the RSRP prediction considering the distribution of real RSRP values.
    \item We propose to use a VAE to learn the environmental auxiliary features from the geographical map generated by a DT. This VAE can be trained offline separately, which does not affect the training efficiency of the likelihood model. The features are used to assist the training of the NN model.
    \item To the best of our knowledge, for the first time, a joint training prototype employing a two-tier neural network for radio coverage prediction is put forward, which benefit from both a digitized simulation model and real data. 
    \item We validated the proposed model using real network data which illustrates superiority over the empirical model.
\end{itemize}

\section{Background}
\label{sec:background}
This section describes the two commonly used RSRP prediction approaches: the empirical-based model and the data-driven approach.
\subsection{Empirical-based model}
Empirical-based models are a set of models summarised from a large amount of measured data in different scenarios. The Log Distance Path Loss (LDPL) propagation model is the most representative one which has been adopted widely \cite{alimpertis2019city}. LDPL treats the power at location $l_j$ as a log-normal random variable, depicts the relationship between the received power and BS to UE distance ${\parallel{l_{BS}-l_{j}}\parallel}_2$, which can be represented by:
\begin{equation}
    P(l_j) = P_0 - 10n_j\log_{10}({\parallel{l_{BS}-l_{j}}\parallel}_2/d_0)+w_j
\end{equation}
where $d_0$ is the close-in reference distance which is determined from measurements close to the transmitter, $n_j$ and $w_j$ are adjustable coefficients determined by the propagation environment \cite{rappaport1996wireless}. After delicate tuning and testing of these parameters, there are several standardised empirical models for propagation modelling that are largely used by academics as well as in industry such as the COST 231 Hata model \cite{7538693}, Okumura model \cite{1622772}, Walfisch-Ikegami model \cite{1232163}, WINNER II Propagation model, etc. However, in the presented work, classical ML techniques and a linear regression model have been employed on the real network data for network radio propagation assessment.

\subsection{Data-driven approach}
The data-driven approach aims to use the historical data of RSRP to analyze the relationship between RSRP and environmental changes over time and space, to achieve fine-grained, site-specific modelling. The intensive development of ML provides a powerful engine for this type of approach. For example, a random forest (RFs) based predictor considering a rich set of features that includes location, time, cell ID, device hardware and other features has been proposed in \cite{alimpertis2019city}. The paper demonstrates the benefits of using fewer measurements and achieving higher accuracy in real-world data sets. Ref \cite{8377405} utilises a Regional Analysis to Infer KPI (RAIK) framework to establish a relationship between geographical data and user data using crowdsourced measurements. A radio wave propagation prediction based on backpropagation (BP) NN and a simplified path loss model is proposed in \cite{wang2020new}. Ref \cite{sohrabi2017construction} proposed a two-step algorithm for RSRP map generation by regression clustering.

However, the challenge for the data-driven RSRP prediction is that there are only a limited number of training features available. In the current work latitude, longitude, altitude, transmission frequency, timestamp, and cell ID are a few of the parameters that were used. The available MDT data does not fully reflect the channel variation. In the NN-based approaches, this limitation will usually result in the underfitting of a NN model. Hence, how to generate more auxiliary features to assist the training process of the NN model and improve the model performance is a significant challenge. Except for the radio features recorded by the BS, some researchers propose extraction and utilisation of environmental features. It is known that geographical statistics influence the signal quality, so the characteristics of the transmission environment, such as the UE distribution and maps, the obstructions in the transmission path, the height of buildings, etc., are integrated with the BS-recorded features. The representative of this method is ray-tracing, which is a popular approximation using geometrical optics and knife-edge diffraction theory \cite{zhang2020cellular}. An ML-based 3D propagation ray-tracing model for the cellular network is studied by \cite{masood2019machine}. Thrane et al. proposed a channel model using deep learning (DL) and a simple path loss model aided satellite images, in which the path loss modelling was also finished by a ray-tracing model \cite{8950164}. Zhang et al. introduced a CNN-based NN to reduce the computational complexity in the ray-tracing model \cite{zhang2020cellular}.
Despite these papers claiming the advantages of less ray-tracing time and the potential in improving path loss prediction, it is far from a complete solution because the proposed NN ray-tracing models are trained in a supervised way, which inevitably need to execute ray-tracing module and collect dataset in advance.
\begin{figure*}[t]   
        \centering   
        \includegraphics[width=1.7\columnwidth]{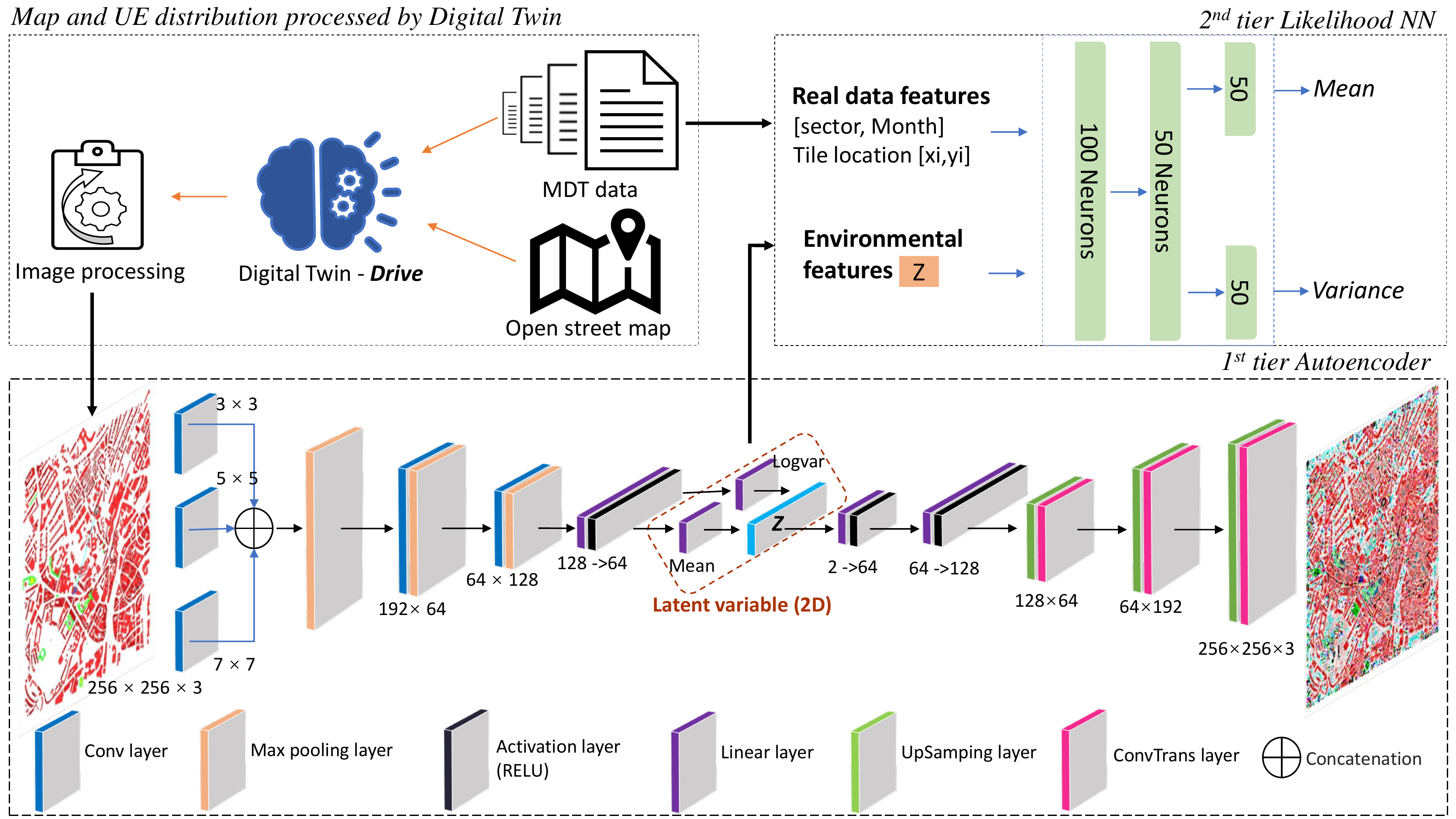} 
            \caption{The proposed two-tier neural network architecture} \label{fig:system architecture}
    \end{figure*} 
    
\subsection{Digital Twin}
\label{sec:DT}
``A digital twin is a digital representation of a physical item or assembly using integrated simulations and service data'' as defined in \cite{erkoyuncu2018digital}. A DT provides high-fidelity representations of all components of the current live mobile network, including service and UE behavioural characteristics \cite{li2021rlops}. With the maturity of image processing technology based on NNs, pure image-based environmental feature extraction schemes are gaining attention. In \cite{thrane2020deep}, the geographical features are processed by the open street map (OSM) and expert knowledge is used jointly to learn a prediction model. Yi et al. put forward an environmental features exploring method by using CNN on the maps of altitude, maps of building height and maps of CI \cite{9351998}. But there are some issues with these methods. For example, the latent features extracted by CNN are sparse, which makes it difficult to identify its effects in training; the CNN-RSRP training is done end-to-end, which makes the model hard to apply to other scenarios. 
In this paper, we utilise a DT, named DRIVE, that was designed by Ioannis et. al. \cite{mavromatis2020drive} to digitalise the BS to UE transmission links in the given scenario. DRIVE is a flexible, modular and city-scale framework aimed at the vehicular and network simulator. It contains three major functionalities -  
\begin{itemize}
    \item It is designed to parse and simplify the OSM and different types of buildings.
    \item The SUMO module is integrated to generate the UE mobility traces according to the road situation from OSM.
    \item It can perform the continuous network simulation according to the UE locations and BS settings.
\end{itemize}
Although the empirical propagation model is adopted in DRIVE, we utilise its OSM processing and the specific UE generation ability that are valuable in modelling the BS to UE relationship geographically.

It is known that the key factors affecting signal transmission exist in the characteristics of the environment, such as how far the signal transmits, and how many reflections, absorption, and scatterings it encountered during this period, even the building materials and vegetation. But these characteristics are difficult to model accurately. Although ray-tracing methods are aiming to restore the transmission path as much as possible, urban-scale ray-tracing is too time-consuming and complicated to be realistic. Moreover, as discussed in \cite{8950164}, simpler images would improve not only the training of the model and the hyperparameter search but also the final performance of the methodology. Therefore, in this paper, we do not seek accurate tracing results. We focus on how to describe the possible impact of the signal transmission path with UE location and environmental information. 

\section{The proposed two-tier NN architecture}
\label{sec:2-tier NN architecture}
In this section, we elaborate on the proposed two-tier NN model with emphasis on NN architecture, training data generation and training scheme. As shown in Figure \ref{fig:system architecture}, the proposed NN model consists of two cascaded NNs. The first-tier is designed as a CNN-based VAE to extract relevant environmental features while the second-tier network is designed as a fully connected network with two heads that outputs the mean and variance of RSRP in a given location. The underlying representation of VAE is $\mathcal{Z}$ - this parameter will assist the training of the two-tier neural network.

\subsection{First tier - Variational Autoencoder}
\subsubsection{Data sampling}

\begin{figure}[t]   
        \centering   
        \includegraphics[width=0.82\columnwidth]{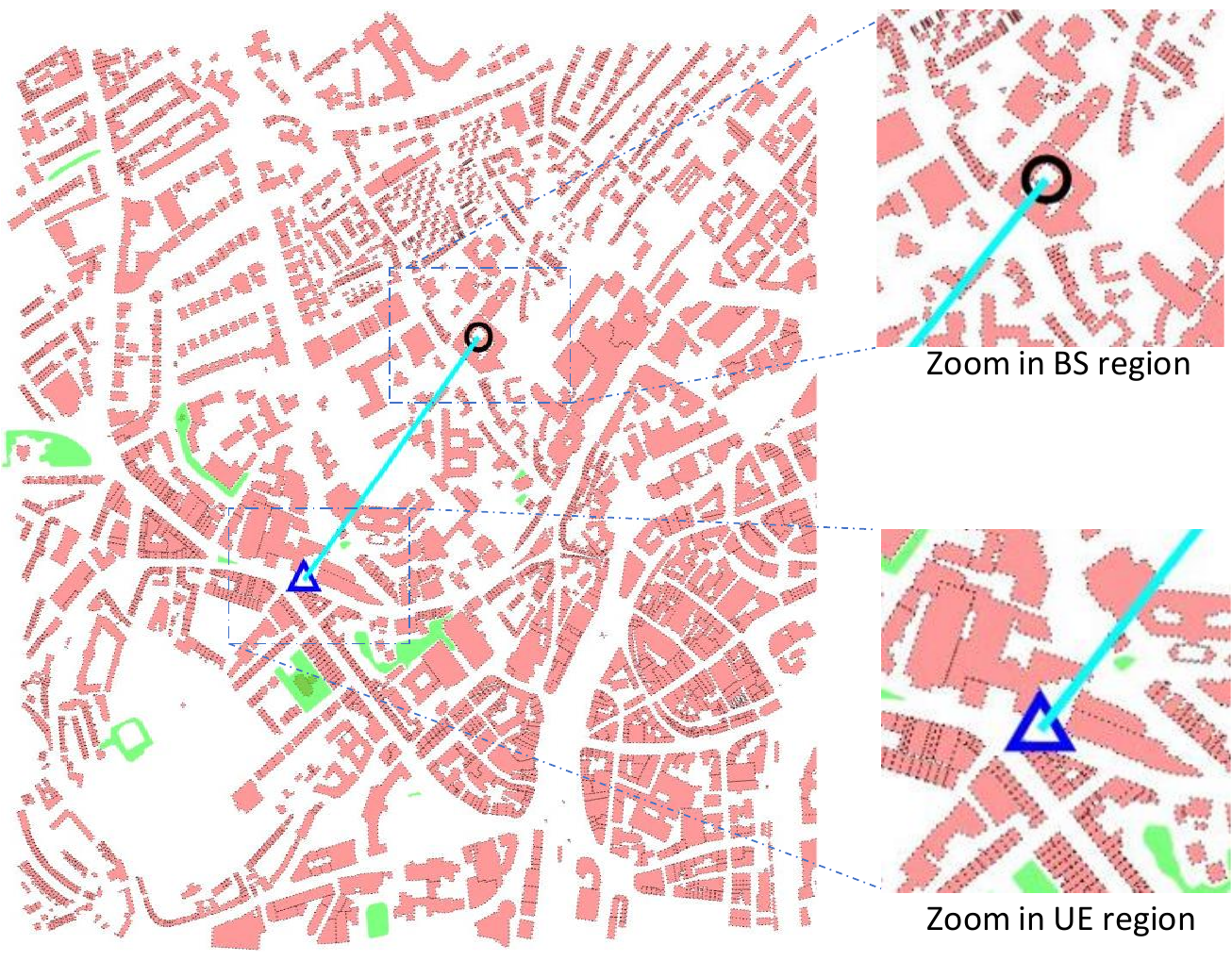}   
            \caption{BS-UE association image generated from DRIVE simulator} \label{fig:UE BS association}
\end{figure} 

As shown in Figure \ref{fig:system architecture}, the OSM corresponding to the site is imported into the DRIVE simulator. Further, the functions provided by the simulator are utilised to simplify the building and road information to understand the building outline of the involved city. In the map processing, the buildings are accounted for as 2D simple polygons as the real data set described in \ref{sec:dataset} doesn't contain accurate altitude information of UEs. Figure \ref{fig:UE BS association} shows an overall association between a UE and a BS. The left side of Figure \ref{fig:UE BS association} shows the processed map, in which red polygons represent typical buildings, and green polygons represent the foliage. BS is represented by a black circle, and UE is marked by a blue triangle. The right side of the figure shows zoomed areas of BS and UE. The connection between the BS and the UE is highlighted by a light blue line. 10000 such top-view geographical images with resolution $256\times256\times3$ are collected to train the VAE, as described below. 
\subsubsection{VAE architecture}
VAE is a framework to learn deep latent-variable models and corresponding posterior inference models using stochastic gradient descent~\cite{kingma2013vae}. It consists of two sections, encoder and decoder, as shown in the lower half of Figure \ref{fig:system architecture}. The encoder, also called the inference model, learns the posterior on the low-dimensional latent space over the input data samples. The decoder is a generative model that learns the joint distribution of the latent variables and input data. In this paper, the architecture of the encoder is in the form of VGG \cite{simonyan2014very}. At the first convolutional layer of the encoder, a set of convolutional filters with sizes of $3\times3$, $5\times5$, $7\times7$ are applied to extract features from different dimensions, respectively, followed by a max-pooling layer. There are two convolutional layers with max-pooling after that. One flattened layer and one fully connected layer with 64 neurons are followed. Next, the output tensor is fed into a two heads output layer with 2 neurons individually. The reparameterized variable $\mathcal{Z}$ will be the input of the decoder section. The decoder owns inverse architecture compared with the encoder.
A dropout with ratio 0.25 was inserted between the flatten layer and the first fully connection layer to enhance the robustness of training. The loss function is defined in Eq.~\eqref{eq:VAE loss}.
\begin{equation}
\mathcal{L}_{VAE}=-\sum_{i}\Big[\mathbb{E}_{z\sim q_\theta(z|x_i)}\big[\log p_\phi(x_i|z)\big]+\mathbb{D}_{KL}\big(q_\theta(z|x_i)||p(z)\big)\Big]
\label{eq:VAE loss}
\end{equation}
where $\theta$ and $\phi$ are the trainable parameters of the encoder NN and the decoder NN, respectively. $q_\theta(z|x_i)$ is the posterior inference from input sample $i$, $p_\phi(x_i|z)$ the generative model given the latent distribution~\cite{odaibo2019tutorial}.
The first term in Eq. \eqref{eq:VAE loss} is the expected data log-likelihood (assuming Gaussian probability density function, maximisation of this term amounts to minimisation of the reconstruction mean squared error), and the second term is the KL divergence between $q_\theta(z|x_i)$ and $p(z)$ which regularises the latent space. The R,G,B channel of each sample $i$ is normalised to the range $[-1,1]$ for training.

\begin{figure}[t]   
        \centering   
        \includegraphics[width=0.82\columnwidth]{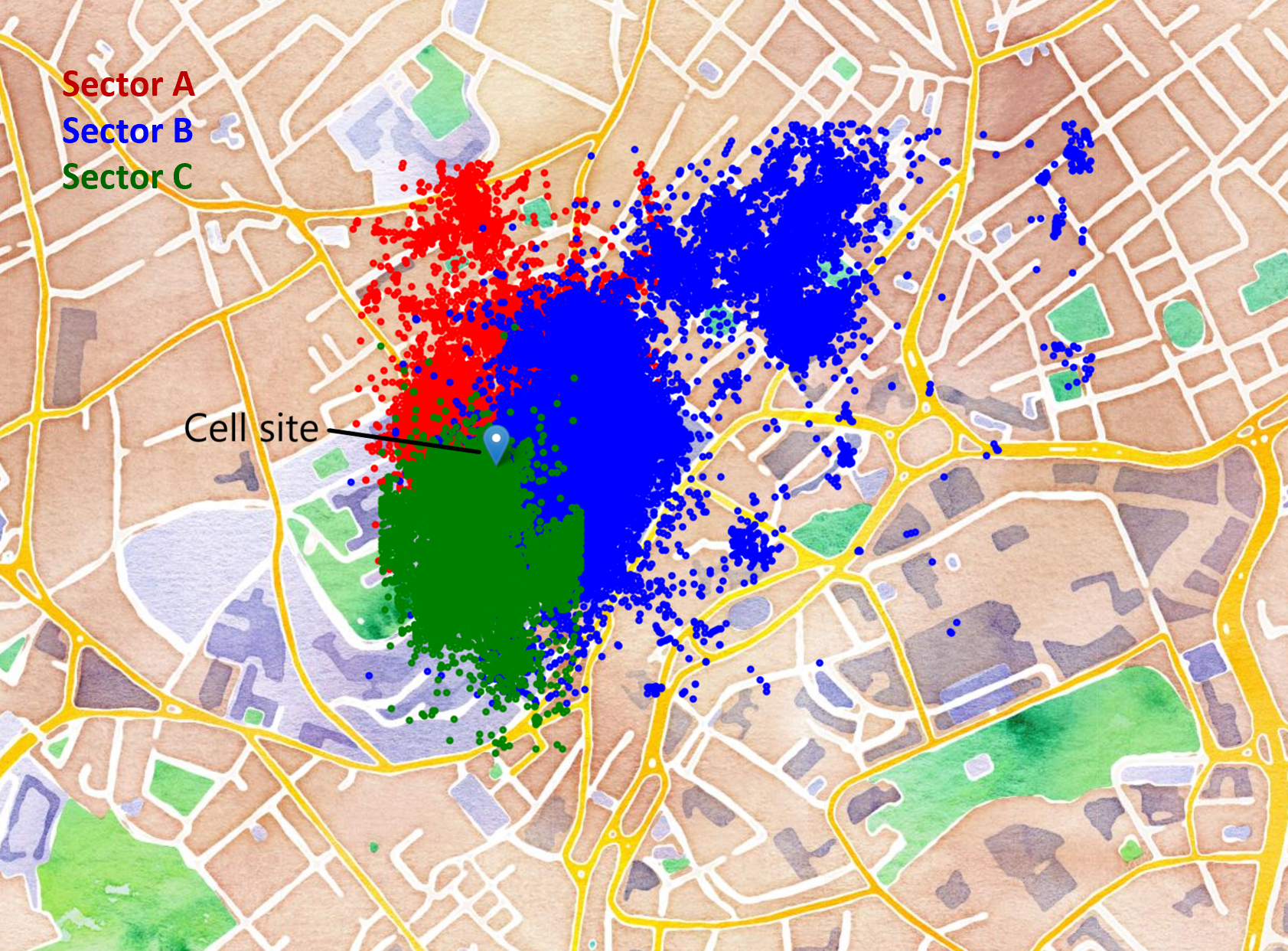}   
            \caption{The UE distribution after removing outliers. The red, green and blue dots indicate UE that are associated with respective sector.} \label{fig:processed_data}
\end{figure} 

\subsection{Second tier - Likelihood Model}
\label{subsec:MLP}
The likelihood model is designed based on an MLP. The detailed architecture is illustrated in the top right side of Figure \ref{fig:system architecture}. The first two layers are with 100 neurons and 50 neurons respectively, and the last layer has two heads with 50 neurons in each head, which output the mean and variance of each bin.
The overall training feature of this model has been formalised in Eq. \eqref{eq:likelihood features}. It is a 6-dimensional vector, where x\_loc and y\_loc represent the $(x,y)$ coordinates. The BS is 3-sectored. A UE belongs to one of the sectors of the BS it is associated with. Month specifies the month that data samples are collected. It is worth noting that the 2-dimensional auxiliary feature generated by encoder section $\mathcal{Z}$ are also fed into the likelihood model.
\begin{equation}
    \mathcal{\vec{F}}_{\text{proposal}}  = (\text{x\_loc},\text{y\_loc},\text{sector},\text{month},\mathcal{Z})^{d=6}
    \label{eq:likelihood features}
\end{equation}

The loss function for training is the Gaussian negative log likelihood loss, which is defined as:
\begin{equation}
    \mathcal{L}_{Likelihood} = \sum_{i}\frac{1}{2}\Big(\big(\frac{(\mu_i-y_i)^2}{\sigma_i}\big)+\ln{{\sigma_i}^2}\Big)
    \label{eq:loss of MLP}
\end{equation}
where $\mu_i$ and $\sigma_i$ are the numerical outputs of the likelihood NN model; $y_i$ is the corresponding label of a sample $i$. Feature normalisation is also adopted for the likelihood model training.
\subsection{Training and Validation}
Experiments are performed using the Intel 2 E5-2640v4 CPU, 2 RTX 2080Ti GPU and 4 × 32G DDR4 SDRAM. The data preprocessing is performed by the CPU whilst the training stage relies on the GPU. The training is based on PyTorch. The training and validation set are divided according to the 80\% and 20\% of the total both for the VAE and likelihood model. The batch size of VAE is $50$ and for the likelihood model is $3000$. Both models use Adam to be an optimiser with the default learning rate.

\section{Real world Datasets and model evaluation}
\label{sec:dataset}



\subsection{Data pre-processing}
The real-world dataset is provided by BT Labs, which records the monthly data of about 16,000 bins served by one BS. Each bin covers a square of $10m\times10m$. A bin may also be referred to as a tile in this work. Each sample of data includes the central coordinate position of the tile and multiple recorded RSRP samples. We chose two datasets with significant seasonality, namely January and August, to evaluate our proposed model. The details of the datasets can be found in the first 3 columns of Table \ref{table:mae_results}. Specifically, the number of samples in each sectors and months. The datasets do not contain any data involving user-specific information.


Due to the uncertainty of the transmission channel, outliers are removed using Hampel’s filter. Here, the median of the whole dataset was first calculated, and then the absolute deviation of each data sample from the median was obtained. Also, the median of the deviations was evaluated. We consider any data point greater than the absolute deviation against the value of $4.5\times$median of the deviations as an outlier. This Hampel's filter is applied on the latitude and then on longitude. The UE distributions of August after removing outliers can be seen in Figure \ref{fig:processed_data}.

\subsection{Evaluation results}

In this section, we provide evaluation results of the proposed two-tier NN model using two real datasets. We evaluate the models in terms of the average RSRP prediction error through a 20-fold cross-validation scheme, and early stopping is adopted in the training process with a stop patience of $8$. The VAE was trained in an offline way. The VAE models with the same parameters were used to assist different likelihood models training under different months. Linear regression and simple MLP techniques are used as the baseline for compassion. The MLP model doesn't contain the environmental extractor and mimics the same architecture in Section \ref{subsec:MLP}. The training feature of this model is defined in equation (\ref{equ:real features}). To keep the consistency of evaluation, the training and validation set divide is the same as in the proposed two-tier NN model, and both models will be reinitialised for each fold. Three sectors involved in the datasets are validated independently.

\begin{equation}
    \mathcal{\vec{F}}_{\text{baseline}}  = (\text{x\_loc},\text{y\_loc},\text{sector},\text{month})^{d=4}
    \label{equ:real features}
\end{equation}
The evaluation criteria is the mean absolute error (MAE) between ground-truth mean value $a_i$ and predicted mean value $\hat{a}_i$. The MAE is calculated as 
\begin{align}
    \text{MAE}=\frac{1}{N}\sum_{i=1}^N{|a_i-\hat{a}_i|}.
\end{align}

\begin{table}[t]
\centering
\caption{Data information and model validation results for different data subset}
\label{table:mae_results}
\begin{tabular}{c|c|c|c|c|c}
\toprule
\multicolumn{3}{c}{Data information} & \multicolumn{3}{|c}{Validation results (in dBm)} \\ \midrule
Month & Sector & Samples & Empirical & MLP & Proposed Model \\ \midrule
\multirow{3}{*}{Jan.} &A &21236 & 7.29 & 6.478 & \textbf{5.840} \\ 
                         &B &53208 & 7.99 & 7.323 & \textbf{6.243} \\ 
                         &C &15172 & 10.71 & 6.758 & \textbf{6.636} \\
\midrule
\multirow{3}{*}{Aug.}  &A &10699 & 8.06 & 7.104 & \textbf{5.941}  \\ 
                         &B &24361 & 10.08 & 9.726 & \textbf{8.623}  \\ 
                         &C &20228 & 8.78 & 7.790 & \textbf{7.012} \\

\bottomrule
\end{tabular}
\end{table}



Table \ref{table:mae_results} presents MAE results of the empirical model, MLP model and proposed two-tier NN. It can be seen that compared with the empirical model our proposed model can improve the prediction accuracy by about 20\%, and the largest increase accrues on the subset January sector C, where the MAE is reduced from 10.71 dBm to 6.758dBm, about 38\%. Meanwhile, compare with the simple MLP model, the prediction accuracy of our proposed model has an improvement by nearly 10\%, and the largest improvement lies in the August sector A, around 16.4\%.

\begin{figure}[t]   
        \centering   
        \includegraphics[width=1\columnwidth]{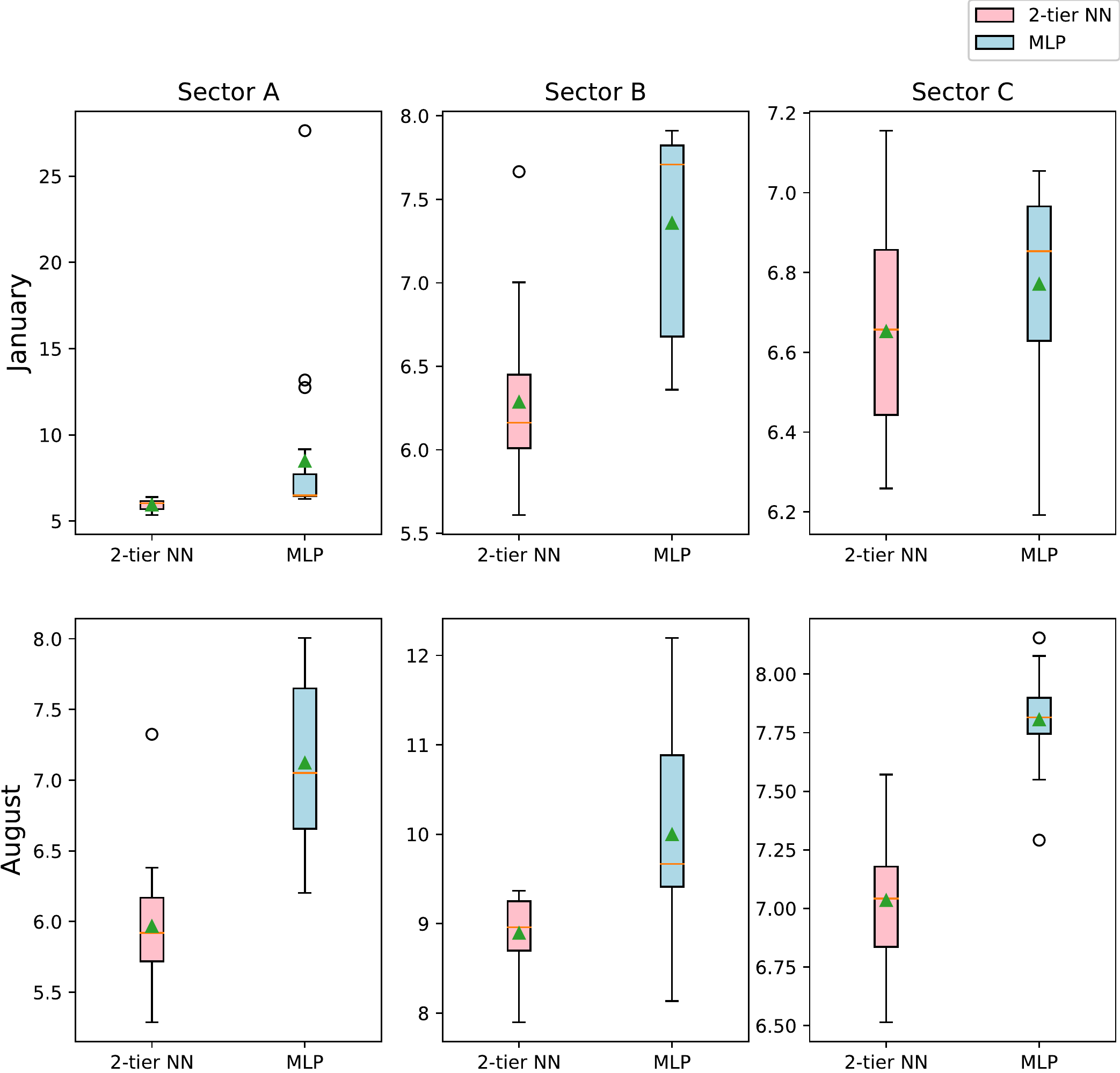}
            \caption{The boxplots of cross-validated MAE for RSRP prediction based on different sectors and months. The boxplots marked red indicate the results of proposed 2-tier NN while the blue boxplots are that of the typical MLP.} \label{fig:boxplot_aug}
\end{figure}

Figure \ref{fig:boxplot_aug} demonstrates more detailed boxplot results, which summarize the distribution characteristics of the MAE on the test set in 20 fold cross-validation for both MLP and the proposed model. It can be seen that, in general, our proposed model can be trained more stable (with fewer outliers) and have a smaller and more concentrated error distribution.


\section{Discussion}
\label{sec:Discussion}
This paper presents a two-tier RSRP prediction model based on OSM processing and demonstrates gains across different real-world datasets. The training of VAE as an environmental information extractor can be separated from the subsequent network, which reflects good model reusability. Since the actual datasets do not contain the altitude information of the UE, so we do not regard the height as an independent one-dimensional feature during map processing and labelling. But the current model can easily implement the above extensions. In addition, the VAE latent vector is equivalent to regularising the training of the subsequent likelihood model, so the length of the latent vector needs to consider the number of real data features, and an excessively long latent vector will suppress the expression of real data features. The likelihood model involved in this paper has the simultaneous output of mean and variance to optimize the loss function shown in Eq. \eqref{eq:loss of MLP}. Since in the current dataset, not all tiles have unified multiple samples recorded and obey the Gaussian distribution. So the output of variance is meaningless for some tiles. Therefore its value can only be used as a reference for partial tiles.
\section{Conclusions}
\label{sec:Conclusion}
In this paper, a novel two-tier NN architecture is proposed to realise the accurate RSRP prediction. The VAE-based environmental feature extractor constitutes the first-tier network which is used to distil the critical information from BS-UE association top-view geographical images, where the image generation is finished in a modified DT (DRIVE) by using OSM of the given area. Meanwhile, the second tier is designed as a likelihood model which takes the outputs of the above extractor and real data features for training. The numerical results evaluated on real-world datasets show the gains of the proposed model in terms of prediction accuracy. The overall accuracy improvement is more than 20\% and around 10\% compared with the empirical and a simple MLP model respectively, and it can reach 38\% and 16.4\% improvement in the best validation case.

\section*{Acknowledgment}
This work was developed within the Innovate UK/CELTIC-NEXT European collaborative project on AIMM (AI-enabled Massive MIMO). This work has also been funded in part by the Next-Generation Converged Digital Infrastructure (NG-CDI) Project, supported by BT and Engineering and Physical Sciences Research Council (EPSRC), Grant ref. EP/R004935/1.
\bibliographystyle{IEEEtran} %
\bibliography{IEEEabrv,references} 

\end{document}